%% file: paper.tex
\documentclass[sigconf,nonacm]{acmart}

\usepackage{multirow}
\copyrightyear{2025}
\acmYear{2025}
\setcopyright{acmlicensed}\acmConference[UbiComp Companion '25]{Companion of the 2025 ACM International Joint Conference on Pervasive and Ubiquitous Computing}{October 12--16, 2025}{Espoo, Finland}
\acmBooktitle{Companion of the 2025 ACM International Joint Conference on Pervasive and Ubiquitous Computing (UbiComp Companion '25), October 12--16, 2025, Espoo, Finland}
\acmDOI{10.1145/3714394.3756153}
\acmISBN{979-8-4007-1477-1/2025/10}
\begin{document}

\title{Enhancing Wearable Tap Water Audio Detection Through Subclass Annotation in the HD-Epic Dataset}

\author{Robin Burchard}
\email{robin.burchard@uni-siegen.de}
\orcid{0000-0002-4130-5287}
\affiliation{%
  \institution{University of Siegen}
  \streetaddress{Hölderlinstr. 3A}
  \city{Siegen}
  \country{Germany}
  \postcode{57076}
}

\author{Kristof Van Laerhoven}
\email{kvl@eti.uni-siegen.de}
\orcid{0000-0001-5296-5347}

\affiliation{%
  \institution{University of Siegen}
  \streetaddress{Hölderlinstr. 3A}
  \city{Siegen}
  \country{Germany}
  \postcode{57076}
}

\renewcommand{\shortauthors}{Burchard et al.}

\begin{abstract}
Wearable human activity recognition has been shown to benefit from the inclusion of acoustic data, as the sounds around a person often contain valuable context. However, due to privacy concerns, it is usually not ethically feasible to record and save microphone data from the device, since the audio could, for instance, also contain private conversations. Rather, the data should be processed locally, which in turn requires processing power and consumes energy on the wearable device. One special use case of contextual information that can be utilized to augment special tasks in human activity recognition is water flow detection, which can, e.g., be used to aid wearable hand washing detection. We created a new label called \textit{tap water} for the recently released HD-Epic data set, creating 717 hand-labeled annotations of tap water flow, based on existing annotations of the \textit{water} class. We analyzed the relation of \textit{tap water} and \textit{water} in the dataset and additionally trained and evaluated two lightweight classifiers to evaluate the newly added label class, showing that the new class can be learned more easily.
\end{abstract}



\maketitle

\section{Introduction}
Human Activity Recognition (HAR) is an increasingly multimodal field \cite{zhangDeepLearningHuman2022,karimHumanActionRecognition2024,wangDeepLearningSensorbased2019}.
In addition to the traditionally used IMUs and RGB-cameras, a multitude of other sensing modalities can be employed to add context and additional information to the existing HAR systems. Such modalities include, but are not limited to, environmental sensors (humidity, pressure, temperature, brightness) \cite{barnaStudyHumanActivity2019, burchardMultimodalAtmosphericSensing2025}, biosignals (e.g., heart rate, electrodermal activity, electromyography), or proximity sensors \cite{fuSensingTechnologyHuman2020}.

Another promising example of an additional modality is sound, which can be recorded with commercially available microphones with small effort and low cost. Acoustic activity recognition and sound event classification are two subproblems one might want to approach when using microphones. Acoustic sensing can be employed to standalone classifiers \cite{hwangEnvironmentalAudioScene2012, liangAudioBasedActivitiesDaily2019}. While RGB(d)-cameras perform worse when there are many occlusions, acoustic and IMU-based models are challenged by ambiguities. Huh et al. describe this fact with the words: ``Some actions are audibly indistinguishable, e.g. ‘wash carrot’ vs ‘wash tomato’, as it is impossible to determine which vegetable is being washed through sound alone.'' \cite{huhEpicSoundsLargeScaleDataset2023,EPICSOUNDS2025}.

Rather than relying on sound alone, recent works propose using sounds as an additional modality to IMUs, RGB-cameras, or other systems \cite{chengAudioSignalProcessing2016, cristinaAudioVideoBasedHuman2024, zhangAudioAdaptiveActivityRecognition2022}. With the added context, systems could become more reliable in picking up unusual instances of specific activities and also become more robust against false positives. 
One example of this is the problem of detecting hand washing among similar activities using IMU sensors: The typical pattern of hand washing often involves rapid and repetitive movements. However, these patterns are not always present for every hand washing instance, and similar patterns can also appear in other contexts, e.g., when brushing one's teeth \cite{burchardWashSpotRealTimeSpotting2022}.
With a microphone, the sound of the activated tap, as well as the water's ``splashing'' could be picked up \cite{taatiWaterFlowDetection2010, zhuangDetectingHandHygienic2023}. By including the information, whether a running tap could be detected nearby, both the rates of false negative and false positive detections could likely be lowered. A similar solution is allegedly used for the Apple Watch's hand-washing timer feature, but the implementation remains closed source \cite{hayesAppleWatchCan2020, appleinc.SetHandwashingApple2024}.

Existing sound datasets often include water-related sounds in one way or another, but the distinction of \textit{tap water} from other water sounds is included in few. However, since many existing datasets already contain the relevant microphone recordings, just no specific annotations, we were able to use a pre-existing dataset for this work, by adding a new label class for \textit{tap water} and annotating it accordingly.

\paragraph{Applications}
The previously mentioned detection of hand washing is only one application of tap water detection, other applications include the monitoring of cooking (e.g., step detection, washing vegetables, rinsing dishes), cleaning, other hygiene routines (e.g., brushing teeth), bathroom usage, and resource usage monitoring (i.e. detecting water wastage). In all these fields, a wearable device could be used to support the user by tracking or habit coaching them, or help them to conserve water. Treating \textit{tap water} as its own separate class has multiple advantages. Tap water is strongly associated with some specific, intentional, high-relevance activities, like the ones mentioned above. In contrast, the general \textit{water} label in audio datasets includes diverse sounds like rain, streams, splashes, gurgling water in pipes, and pouring. These water-related sounds vary greatly in frequency, duration, and intensity. \textit{Tap water} is more homogeneous, and therefore easier to learn for machine learning models, as we show in this work.

An additional constraint of recording audio data is protecting all users' and non-users' privacy. Since microphone recordings could accidentally include private conversations, clues about the users' whereabouts, or other personal information, they have to be handled with a high level of care. Ideally, audio recordings should never leave the recording device and should be immediately removed after processing. This would ensure the highest possible protection of the users' privacy. To make this possible, classifiers should be small enough and require little complexity, as mobile devices need to be powerful enough to run the required model inference. Focusing on the task of tap water detection, we propose to train models using our new annotations, which we created specifically for this task, on the existing, but also very recently released, HD-Epic dataset. 

The main goal of this work was to create a precisely annotated dataset for the detection of running tap water from microphone data. The dataset can be employed for automated tap water detection on wearable devices, which in turn has a multitude of applications.

Our main contributions in this paper are as follows:
\begin{enumerate}
    \item We motivate why on-device \textit{tap water} detection is a crucial task in activity recognition.
    \item We hand-labeled a precise, new class in the HD-Epic dataset, resulting in 717 instances of \textit{tap water} audio events.
    \item We show how this new \textit{tap water} class compares to the existing \textit{water} class, demonstrating how HD-Epic's sound annotations can be used for audio-based activity recognition 
    \item We created a pre-trained CNN ExecuTorch model to detect \textit{tap water} for deployment on wearables 
\end{enumerate}

\paragraph{Ensuring reproducibility:} All code, data, annotations, and other information needed to reproduce the results of this work can be found online, in our GitHub repository\footnote{\label{fn:github}\url{https://github.com/RBurchard/hd-epic-tap}}.

\section{Related Work}
\input{tab/datasets}
Several annotated sound datasets containing \textit{water} or \textit{tap water} have been published in the past. 
Table \ref{tab:datasets} allows a quick comparison between the available datasets and our work. Notably, Perett et al. have published HD-Epic, which is used as the basis of this work \cite{perrettHDEPICHighlyDetailedEgocentric2025}. Before that, an earlier version called Epic Sounds \cite{huhEpicSoundsLargeScaleDataset2023,EPICSOUNDS2025} was released as a derived work of the EPIC-KITCHENS-100 dataset \cite{damenRescalingEgocentricVision2022}. These two datasets contain \textit{water} as a class, but they avoid distinguishing it further into different sub-classes of water-related sounds. The ESC-50 dataset for environmental sound classification (ESC) \cite{piczakESCDatasetEnvironmental2015} by Piczak et al. contains a collection of water-related sound categories like \textit{rain}, \textit{sea waves}, \textit{water drops}, \textit{pouring water}, \textit{toilet flush}, and \textit{thunderstorm}. However, the dataset does not contain \textit{tap water} as a class. AudioSet by Gemmeke et al. \cite{gemmekeAudioSetOntology2017} contains 10s audio sample clips drawn from YouTube videos, which were categorized into 527 sound event classes. The dataset includes 2442 of these 10s segments, in which \textit{tap water} is present. However, the label for each 10s audio clip only provides the information that at some point in the clip \textit{tap water} is present, and only $1\,Hz$ embeddings of the audio data are published. Lastly, Fonseca et al. published the FSD50K dataset \cite{fonsecaFSD50KOpenDataset2022}, which contains 51 thousand audio clips from Freesound, which were manually labeled into a selection of 200 classes of AudioSet's ontology. This dataset contains 458 \textit{tap water} clips. Due to the samples being collected from Freesound, different licenses can apply, and the sounds in the FSD50K dataset are usually very specific and distinct audio clips of just one annotated class, without background noise.

To conclude, each of the listed audio datasets has important limitations. Whether due to restricted access, the absence of precise timestamp annotations, or a lack of realistic background sounds that reflect real-world conditions.

In the domain of audio event classification, many different audio event recognition systems exist. Like in many machine learning domains, feature extraction is an important part of such a system's pipeline, and features usually include Fourier transforms, cepstral features, spectrograms, wavelet transforms, and statistical properties of the audio signal \cite{prashanthReviewDeepLearning2024,cristinaAudioVideoBasedHuman2024}. For training classic machine learning models, such as support vector machines or random forests, spectral and statistical features are used \cite{chuEnvironmentalSoundRecognition2009,piresUserEnvironmentDetection2017}. To train deep learning models, existing research often relies on log-mel spectrograms as the input to a convolutional neural network (CNN) \cite{hyunSoundEventDetectionWaterUsage2023,yuhRealTimeSoundEvent2021}. 

\section{Methods}
In this section, we report how we created the new label class \textit{tap water} on a pre-existing dataset. Furthermore, we describe our findings from the labeling process and the validation of the newly created annotations. Lastly, we outline the creation of an automated detection system for \textit{tap water} detection on mobile devices using lightweight machine learning algorithms.

\subsection{Dataset}
We used the publicly available HD-EPIC dataset \cite{perrettHDEPICHighlyDetailedEgocentric2025} with an adapted label class. The original dataset contains activities from nine participants, together with video and audio data, recorded at the participants' homes. We focus solely on the audio data, for which the dataset contains 50,968 audio annotations of 44 classes, such as \textit{click}, \textit{rustle}, or \textit{water}. We chose the EPIC-HD dataset because it offers high audio quality (48\,kHz sampling rate) and has a high quality of annotations. The dataset is also recorded in-the-wild, in nine different indoor environments, which helps to avoid overfitting to a specific environment. However, none of the classes in the existing annotations align perfectly with our focus on acoustic tap water flow detection. Thus, we used the \textit{water} and \textit{pour} classes as a starting point to annotate a new class, \textit{tap water}.

\subsection{Label creation for the new class \textit{tap water}}
\label{sec:labelcreation}
A single person, the first author of this publication, manually annotated the entire dataset of nine participants. To do so, we did not go through all video recordings again at their entire length, but rather focused on improving the existing labels of \textit{water} and \textit{pour} for our task. This was based on the idea that \textit{tap water} should be a subset of \textit{water} and \textit{pour}. We include \textit{pour}, because the word \textit{pour} can also imply water and tap water flowing, depending on interpretations of natural language. However, we found that the original annotations did not contain any tap water under \textit{pour} labels. During label creation, we gained some valuable insights into the audio aspect of HD-Epic, which we also describe in this subsection. 

\subsubsection{Description of labeling criteria and initial results}
\label{sec:labelcritres}
The original HD-Epic dataset contains 2974 \textit{water} labels, and 171 instances of \textit{pour}. After analyzing the durations of the separate labels, we found that roughly two-thirds of the labels were so short, that they only made up for around $11\,\%$ of the total \textit{water} duration. We argue that these short \textit{water} labels are unlikely to contain \textit{tap water} annotations, or even if they do, they are so short that they would have little positive influence on the training and evaluation of a classification model. Therefore, we decided to only relabel labels with a duration of three seconds or more. This cut-off meant that we relabeled only slightly more than one-third of the instances of \textit{water}, but we could still cover almost $90\,\%$ of the time that was labeled as an instance of this class. Out of the initial 2974 \textit{water} and 171 \textit{pour} annotations, we investigated 1058 \textit{water} and 40 \textit{pour} annotations that were longer than three seconds.

We created a tool to jump to \textit{water} and \textit{pour} labels in the recordings, and then, together with the videos, inspected and annotated all occurrences of tap water being audible in the dataset. In the HD-Epic dataset, \textit{water} refers not only to taps running, but to all water-related sounds, including but not limited to the sound of single water drops, drain pipes gurgling, water splashing, and sometimes even the stirring of a pot. Therefore, to obtain \textit{tap water} annotations, we hat to many \textit{water} annotations had to be ignored, shortened, or split up into multiple parts.

After manually labeling the videos of all 9 participants, we received a total of 717 new annotations for \textit{tap water}, with a total duration of 9595 seconds. Leaving out newly created labels that were shorter than 3 seconds, 9439 seconds of labels were created. In total, the ratio of the durations of \textit{tap water} labels compared to \textit{water} labels is $62\,\%$, which means that while \textit{tap water} is the largest subset of the \textit{water} class, it is also significantly smaller. Thus, we argue that it should be treated as a different class than \textit{water}, to allow for higher granularity and better usability for augmenting specific tasks, like hand washing detection.

\input{tab/durations}

Table \ref{tab:durations} shows the durations and ratios of annotated times for the \textit{water} and \textit{tap water} label classes. In total, HD-Epic contains 16981 seconds of \textit{water}, out of which 15182 seconds were used by us for analysis and creating the new label class \textit{tap water}. 

\subsubsection{HD-Epic Dataset Findings}
Overall, the quality of sound labels in the HD-Epic dataset is high. Whilst investigating and augmenting the originally contained audio labels, we obtained some learnings regarding these original audio labels, mostly related to the water class.
\paragraph{Miss-classifications}
There are many sounds present in a kitchen environment that can sound similar to water, and that were therefore occasionally incorrectly labeled as \textit{water} in the dataset. Some examples include stirring a pot, whisking eggs, or the sound of sizzling pans in the background. Additionally, plating non-liquid food from a pan, moving food into a container, or an instance of \textit{pour} without water present were annotated as \textit{water}. Additionally, we found \textit{pour} annotations when a participant was frothing milk in one case, and handling a plastic bag in another case.

\paragraph{Inconsistencies}
We found some inconsistencies in the labels, where similar sounds were treated differently during the label creation. As an example, single drops of water in the background were sometimes labeled and sometimes missed throughout the different recordings and participants. The same problem occurred with water in the background in general. For participant P01, water in the drain and audible pipes were annotated, but not for most of P02's recordings. In some cases, pouring water was annotated as \textit{pour}, and in other cases it was annotated as \textit{water}. Sometimes, multiple water labels were overlapping or contained inside another.

\paragraph{Diversity of water sounds}
The \textit{water} class contains a multitude of different, water-related sounds. The sound of a running tap usually comes with a distinct ``noise'', whereas other water-related sounds can be less specific. Single water droplets make a distinct, short sound, whilst pipes ``gurgling'' in the background produce a completely different sound. Between the sounds of pouring water in a cup, ``splattering'' water in a sink, and a pressure cooker releasing steam, there is a lot of auditory diversity. Although we fully agree that all these sounds should be grouped under the \textit{water} label, we argue that it is naturally difficult to train a classification model for such a diverse class of many-shaped audio labels. 
Therefore, we argue that the less diverse and more consistent class \textit{tap water} is a valuable addition to the HD-Epic dataset.

\subsubsection{Auditory challenges with the new labels}
\label{sec:auditory_challenges}
In the cases of some of the newly created \textit{tap water} audio annotations, issues can be predicted from the characteristics of the contained sounds. Participant P02, e.g., was using a very silent tap, with water sometimes running in the background, far away from the used microphones. Additionally, P02 and P04 sometimes only turned on their taps to a slight drizzle, which was barely audible, and ran without the often-present ``water flow noise''.

In other cases, loud and similar-to-water noises were in the background. For P03 and P06, extractor hoods were sometimes running in the background, making the tap hard to hear.

\subsection{Analysis and validation of the new labels}
\label{sec:labelvalidation}

\paragraph{Intersection over Union}
To better understand the similarity of the \textit{tap water} class to the \textit{water} class, we report the intersection over union (IoU) of the two classes, calculated and aggregated over each \textit{water} annotation (see Eq. \ref{eg:iou}). The calculation of the IoU is done by merging all overlapping \textit{water} and \textit{tap water} regions and then aggregating their IoUs over the entire dataset. 
\begin{align}
    \label{eg:iou}
    \text{IoU} = \frac{\left| \left(\texttt{tap water} \cap \texttt{water} \right)\right|}{\left| \left(\texttt{tap water} \cup \texttt{water} \right)\right|}
\end{align}

Table \ref{tab:iou_per_part} shows the IoU per participant and over the entire dataset. The IoU spans from $0.364$ to $0.916$ with a total IoU of $0.616$ calculated on the dataset. 
\input{tab/iou_per_part}

\paragraph{Coverage of \textit{tap water} by \textit{water}}
We also report the coverage in Table \ref{tab:iou_per_part}, which is defined as the ratio of containment of \textit{tap water}, i.e., the ratio of \textit{tap water} that is contained inside \textit{water} annotations (see Eq. \ref{eq:coverage}).
\begin{align}
\label{eq:coverage}
\text{coverage} = \frac{\left| \left(\texttt{tap water} \cap \texttt{water} \right)\right|}{\left|\left(\texttt{tap water} \right)\right|}
\end{align}
The coverage is in the range of $93\,\%$ to $99.9\,\%$ across all participants, and has a value of $97.8\,\%$ for the entire dataset. The coverage value is high, as we only annotated near the \textit{water} labels of the original dataset. However, the value also supports the initial assumption that \textit{tap water} is a subclass of \textit{water}, as it is mostly covered by \textit{water} annotations.

To conclude this subsection, our initial validation of the \textit{tap water} labels shows that we successfully annotated a subclass of \textit{water}, and that \textit{tap water} makes up for around $63\,\%$ of all water-related sound in HD-Epic's \textit{water} annotations.

\subsection{Machine Learning Experiments}

To validate the audio detection of \textit{tap water}, we trained machine learning classifiers, and additionally compared the results with the same classifiers trained on \textit{water}.

\subsubsection{Classification tasks}
We define the task of classifying \textit{water} or \textit{tap water} as the positive class against all other classes as the Null class. We report the results both for a $70-30\,\%$ train-test-split over all participants (Task A), as well as for a more realistic leave-one-participant-out (LOPO) paradigm (Task B). The train-test-split shows how well the classifiers perform for each of the positive classes under conditions where individual-specific patterns may be learned during training.
The LOPO split yields a more realistic estimate of real-world performance: Since the participant and environment used for evaluation are not shown to the model during training, the performance is indicative of the real-world performance we could expect for unseen environments. 

\subsubsection{Features}
\label{sec:features}
Similar to previous publications, we used frequency-domain feature vectors. We focused on the Mel-Frequency Cepstral Coefficients (MFCCs), spectral features such as centroid, bandwidth, contrast, rolloff, cover (\cite{guyotWaterFlowDetection2012}), as well as the chromagram, zero crossing rate, and the root-mean-square energy (RMSE) of the audio signal. We calculated each of these for 2-second-long sliding windows without overlap. The features were calculated using the implementations contained by the ``librosa'' python library, and then aggregated for each window, using the mean, sd, min, max, and median as aggregation functions. In total, this left us with a 41-dimensional feature vector.

\begin{figure}[h!]
    \centering\vspace{-3mm}
    \includegraphics[trim={0.4cm 0 1cm 0},clip,width=1\linewidth]{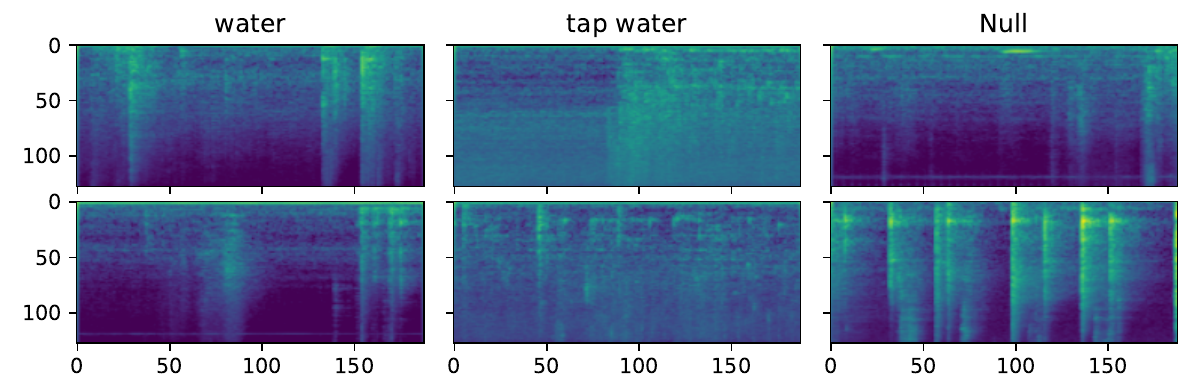}\vspace{-3mm}
    \caption{Two examples per class of the log-mel spectrograms, used as input for the CNNs. The spectrograms for the \textit{water} class are chosen as examples that are not also in \textit{tap water}.}\vspace{-5mm}
    \label{fig:spectrograms}
\end{figure}

In line with the literature, we calculated log-mel spectrograms for the use in a neural network \cite{yuhRealTimeSoundEvent2021, seoConvolutionalNeuralNetworks2022}. Example spectrograms are shown in Figure \ref{fig:spectrograms}, where the intervals chosen show the broad spectrum of ``noise'' that we can expect from tap water sounds.

\subsubsection{Classification models}
 For our intended use of wearable \textit{tap water} detection, models must be small and computationally simple enough to run on mobile CPUs, i.e. we have to refrain from using large transformer models. Since the scope of this work is to show the usefulness of the newly added class, and not to achieve the highest possible results, we stuck to two simple but appropriate classification models.

 We used a random forest (RF) classifier and trained it on the features described in \ref{sec:features}. We used the standard settings of sklearn's implementation, except ``class\_weight'' which we set to ``balanced'' to take the skewed distribution into account, and ``n\_estimators'' which we set to 200. 
 
Similar to the literature \cite{yuhRealTimeSoundEvent2021}, we trained a CNN on the log-mel spectrograms as described in \ref{sec:features}. The CNN consisted of five 2D-convolutional layers, each followed by a 2x2 max-pooling layer. Two fully connected layers form the classification head of the network architecture. We trained the CNN model using the ADAM optimizer and weighted binary cross-entropy as the loss function.

Since the dataset is highly imbalanced for our target classes, we also report a baseline performance, i.e. the best-performing ``dummy classifier'', w.r.t. the F1 score. In our case, the ``uniform'' prediction strategy yielded the best baseline performance throughout all experiments.

\subsubsection{Reported metrics}
We report the accuracy, as well as F1 score, precision, and recall for all tasks, models, and target classes. Since \textit{water} and \textit{tap water} only make up for less than $10\,\%$ of the dataset, accuracy alone is not an expressive metric. Hence, we also report the F1 score and its components, precision and recall, as they provide more information about the actual predictive results. Also, to make the comparison of the performance for the two target classes easier, we calculate the ratio between the classifier performance and the baseline performance for the F1 score metric (R. to Bln.). This performance ratio better indicates the relative performance of the classifier w.r.t. to the general difficulty of the task.

\subsubsection{ExecuTorch runtime model}
To share the models created in this work, we have trained one CNN model on all available data and made it available pre-trained for ExecuTorch, PyTorch's mobile deployment solution. The model can be used to detect \textit{tap water}. As input, it requires windows of computed log-mel spectrograms, as described above. For further details, we refer 
to our 
repository.

\section{Results} 
\label{sec:results}
\begin{figure*}[h!]
    \centering\vspace{-5mm}
    \includegraphics[width=\linewidth]{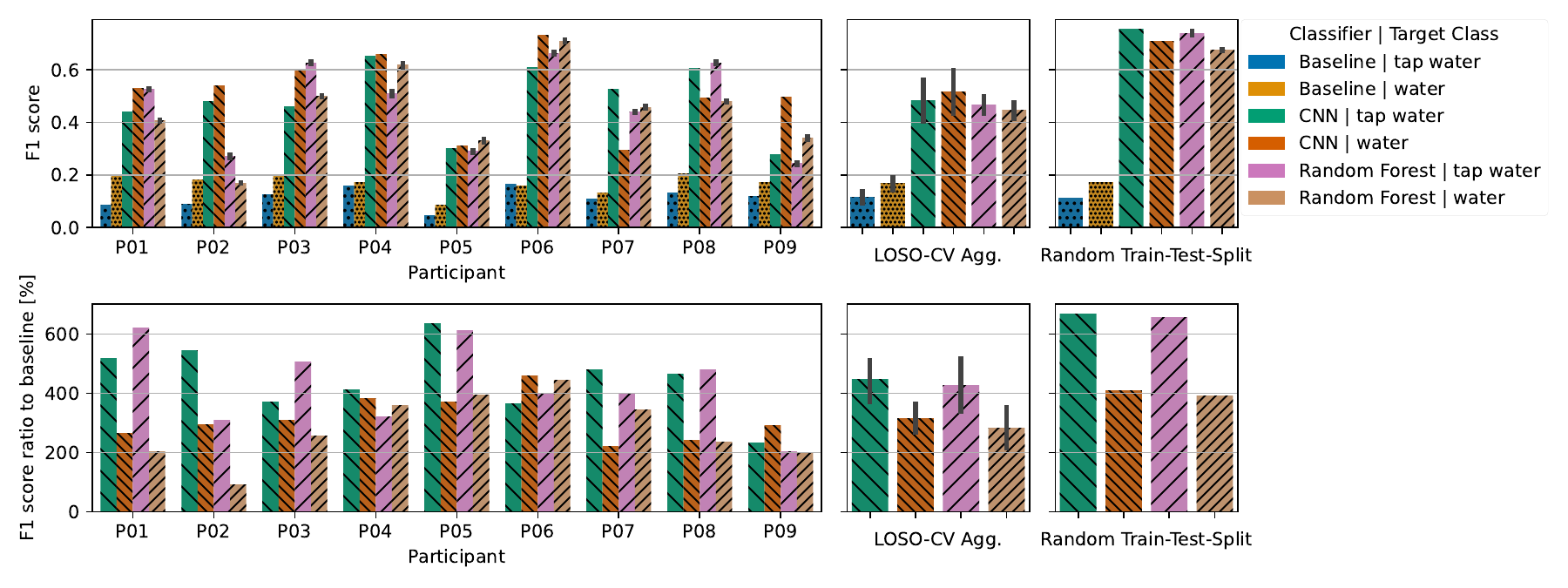}\vspace{-4mm}
    \caption{Results of classifier training and evaluation for Task A and Task B. F1 score results (upper row) are displayed for the leave-one-participant-out cross-validation, per participant (and thus per environment, left) and aggregated (center), as well as for Task A (random train-test-split, right). The baselines are consistently outperformed by both CNN and Random Forest, for both target classes. We also show the ratio of the F1 score to the respective baseline F1 score (lower row), for the same splitting modalities (left: Task B and right: Task A). Classifiers for \textit{tap water} mostly achieve higher ratios than classifiers for \textit{water}.}\vspace{-2mm}
    \label{fig:f1_score_res}
\end{figure*}

\input{tab/results_ml}
Table \ref{tab:res_loso} and Figure \ref{fig:f1_score_res} show the results for Task A and Task B. For Task B, the table only reports aggregated results, while Figure \ref{fig:f1_score_res} includes the results per participant. Overall, the trained classifiers' prediction performance on the \textit{tap water} class is higher than the performance on the \textit{water} class. The low baseline F1 scores of $0.11$ for \textit{tap water} and $0.17$ for \textit{water} are indicative of the dataset's natural imbalance. These baseline results show that it is very difficult to achieve good F1 scores because the class distribution is strongly imbalanced.

For Task A, the Random Forest performs slightly better than the CNN, and for both classifier types, the task of classifying \textit{tap water} against all other classes yields the highest performance, in terms of F1 score, Accuracy, and Precision. The CNN consistently outperforms the Random Forest with its Recall results. The highest F1 scores reached are $0.75$ on \textit{tap water} and $0.71$ on \textit{water} (both CNN).
Due to the distribution, reaching the same F1 score is more difficult for the \textit{tap water} class. However, the general results with a higher F1 score for \textit{tap water} in Task A support our expectation that the \textit{tap water} class is more homogeneous and therefore easier to learn and predict.
This becomes even more apparent when we consider the ratio to the baseline (R. to Bln.) results. For the \textit{tap water} class, the classifiers outperform the baseline by $668\,\%$ (CNN) and $656\,\%$ (RF), while for \textit{water} the ratios are only $410\,\%$ (CNN) and $391\,\%$ (RF).

The performance for the LOSO-CV split of Task B differs strongly from participant to participant. While the baseline results are relatively similar for most participants, they are lower for P05 for both classes, (F1: \textit{water}: $0.08$, \textit{tap water}: $0.05$). P05 is also the participant for which the classifiers achieve the worst cross-validation F1 scores, all around $0.3$. The generalization performance to the unseen participants' data also varies per classifier, i.e., sometimes the CNN has a slight edge over the Random Forest, and vice versa. The best performances are reached for P06, where the CNN achieves an F1 score of $0.73$ for \textit{water} and the Random Forest reaches $0.66$ for \textit{tap water}. On average, the CNN generalizes slightly better on both \textit{tap water} (CNN: $0.48$, RF: $0.47$) and \textit{water} (CNN: $0.52$, RF: $0.45$). The results for all participants are shown in Figure \ref{fig:f1_score_res}.
While the results for the F1 score are similar across classifiers and target class, the ratio to the baseline is still better for \textit{tap water} with a ratio of $447\,\%$ (CNN) compared to $315\,\%$ for \textit{water} (also CNN). A Shapiro-Wilk test (stat $= 0.934$, $p = 0.226$) indicated normality. Both a paired $t$-test ($p = 0.0005$) and Wilcoxon test ($p = 0.0019$) showed that \textit{tap water} significantly outperformed \textit{water} w.r.t. the ratio to the baseline.

Notably, the RF achieves higher accuracy and precision results throughout the experiments, and, in turn, the CNN achieves higher recall scores.

\section{Discussion and Future Work}
We discussed that \textit{tap water} detection has direct applications in smart home systems, hygiene monitoring (e.g., handwashing), and resource conservation. Especially for HAR tasks, the general presence of \textit{water} sounds contains less contextual information than \textit{tap water} being audible.

Through the statistics and validation analysis presented in \ref{sec:labelcritres} and \ref{sec:labelvalidation}, it becomes apparent that \textit{tap water} is mostly a subclass of \textit{water} in the HD-Epic dataset. Its duration in the dataset makes up for around $60\,\%$ of the duration of all \textit{water}. While \textit{tap water} can also represent diverse sounds, depending on the tap used and the water flow intensity, it appears to be more homogeneous than \textit{water}. \textit{Water} represents a superclass of multiple liquid-related sounds that are related, but oftentimes not specific enough for possible use cases, especially in HAR.

The machine learning results presented in Section \ref{sec:results} also hint at \textit{tap water} being easier to learn by classifiers. Especially when compared to the baseline performance, the resulting F1 scores for \textit{tap water} are consistently higher than the F1 scores for \textit{water}. We used lightweight classifiers and did not extensively optimize the model training. In future work, we want to analyze failure cases and relate them to the challenges listed in Section~\ref{sec:auditory_challenges}, searching for potential solutions. A higher raw performance could likely be achieved using state-of-the-art edge-optimized classifiers. In a future application, prediction post-processing, e.g., smoothing, could be used to  improve the detection results, as the smoothing would likely be able to filter out outliers in the prediction.
\section{Conclusions}
This work describes how we created and evaluated a publicly available dataset of precise \textit{tap water} audio annotations in real-world kitchen environments. We based our work on the HD-Epic dataset, which provided a well-annotated basis with a generic \textit{water} class.

We systematically evaluated how the \textit{tap water} audio annotation class is mainly a subclass of \textit{water}, and argued why it is useful to consider it separately. Applications for the newly created labels can be found in a multitude of tasks, such as hand washing detection, hygiene monitoring and compliance, or resource usage surveillance.

By reporting initial classification performance results, we show that despite it representing a less frequent class in the dataset, training classifiers on the more uniform \textit{tap water} is not more difficult than on the \textit{water} class. Especially when compared to the baseline performance, models trained to detect \textit{tap water} achieve higher results.

In this work, we additionally discussed multiple applications of \textit{tap water} detection in acoustic sensing. 

We conclude that the created 717 labels provide a valuable contribution to the audio labels of HD-Epic and can, in future work, be used to train off-line classifiers for \textit{tap water} detection. All annotations, code for generating the results, and the pre-trained model are available in our GitHub repository\footref{fn:github}.

\balance


\bibliographystyle{ACM-Reference-Format}
\bibliography{citations}

\end{document}

%% file: tab/datasets.tex
\begin{table*}[t]
\centering
\vspace{-2mm}
  \caption{Audio datasets containing water-related annotations. The table shows the duration of contained \textit{tap water} annotations, the datasets' sampling frequencies, whether the dataset is open (raw data freely available), the source of the audio data, and whether the dataset is labeled accurately with start/end times for each sound (precise labels). We created 157 minutes of precise annotations on the existing HD-Epic dataset. AudioSet provides its data as 128-dimensional audio embeddings extracted at 1 Hz rather than raw audio.}\vspace{-4mm}
  \label{tab:datasets}
\begin{tabular}{l|l|l|l|l|l|l}
\hline
dataset     & tap water duration (min.) & add. water classes         & freq. (kHz) & open & source & precise labels \\ \hline
Epic Sounds & None                                 & water / pour                        & 24                       & x                  & EPIC-Kitchens-100 & x                 \\
HD-Epic     & None (orig.) / 157 (ours)            & water / pour                        & 48                       & x                  & HD-Epic           & x                 \\
AudioSet    & 407, in 10s segments                 & many, rain, streams, etc. & 1 Hz (embeddings)        & -                  & YouTube           & -                 \\
FSD50K      & $\sim$53, in 458 clips               & waterfall, waves, etc.              & variable, 44.1+          & x                  & Freesound         & -                 \\ 
ESC-50      & None               & rain, waves, etc.              &  44.1          & x                  & Freesound         & -                 \\ \hline
\end{tabular}
\end{table*}

%% file: tab/durations.tex
\begin{table*}[t]
  \centering
  \vspace{-2mm}
  \caption{Durations of \textit{water} and \textit{tap water} labels in the dataset, per participant and aggregated. The table splits the durations up into all labels and labels lasting three seconds or longer. The ratio of the remaining labels, if we remove labels shorter than three seconds, is shown in the third column from the right. On aggregate, 89\% of the duration of \textit{water} annotations are kept, and 98\% of tap water annotations. The ratio between tap water and water annotations is shown in the rightmost columns.}\vspace{-4mm}
  \label{tab:durations}

\begin{tabular}{l|rr|rr|rr|c|c}
\toprule
 & \multicolumn{2}{c|}{duration (s)} & \multicolumn{2}{c|}{duration (s) >= 3s} & \multicolumn{2}{c|}{ratio >= 3s} & ratio & ratio >= 3s \\
participant id & tap water & water & tap water & water & water & tap water  & tap water / water & tap water / water \\
\hline
P01 & 895.38 & 2737.15 & 832.48 & 2491.81 & 0.91 & 0.93 & 0.33 & 0.33 \\
P02 & 765.34 & 2189.15 & 752.25 & 1758.26 & 0.80 & 0.98 & 0.35 & 0.43 \\
P03 & 1821.17 & 3261.02 & 1787.52 & 3025.74 & 0.93 & 0.98 & 0.56 & 0.59 \\
P04 & 1474.32 & 1872.03 & 1457.36 & 1732.96 & 0.93 & 0.99 & 0.79 & 0.84 \\
P05 & 297.81 & 541.46 & 294.93 & 427.84 & 0.79 & 0.99 & 0.55 & 0.69 \\
P06 & 1432.18 & 1588.87 & 1429.38 & 1458.60 & 0.92 & 1.00 & 0.90 & 0.98 \\
P07 & 789.91 & 944.50 & 789.91 & 844.36 & 0.89 & 1.00 & 0.84 & 0.94 \\
P08 & 990.47 & 1838.34 & 984.14 & 1754.68 & 0.95 & 0.99 & 0.54 & 0.56 \\
P09 & 1128.00 & 2008.62 & 1111.22 & 1687.98 & 0.84 & 0.99 & 0.56 & 0.66 \\
\hline
Agg. & 9594.57 & 16981.14 & 9439.19 & 15182.24 & 0.89 & 0.98 & 0.57 & 0.62 \\
\bottomrule
\end{tabular}
\end{table*}

%% file: tab/iou_per_part.tex
\begin{table*}[t]
  \centering
  \vspace{-2mm}
  \caption{IoU: Intersection over Union between the original \textit{water} annotations and the newly created \textit{tap water} class. Coverage: Ratio of containment of \textit{tap water} annotations inside \textit{water}. }\vspace{-4mm}
  \label{tab:iou_per_part}

\begin{tabular}{l|rrrrrrrrr|r}
\toprule
participant & P01 & P02 & P03 & P04 & P05 & P06 & P07 & P08 & P09 & All\\
\hline
IoU & 0.364 & 0.429 & 0.614 & 0.842 & 0.651 & 0.871 & 0.916 & 0.542 & 0.618 & 0.616 \\
coverage & 0.999 & 0.995 & 0.991 & 0.983 & 0.976 & 0.927 & 0.993 & 0.984 & 0.968 & 0.978 \\
\bottomrule
\end{tabular}

\end{table*}

%% file: tab/results_ml.tex
\begin{table*}[t]
  \centering\vspace{-2mm}
  \caption{Results of Task A: Random Train-Test-Split and Task B: Leave-One-Participant-Out Cross-Validation.}\vspace{-4mm}
  \label{tab:res_loso}

\begin{tabular}{l|l|c|c|c|c|c|c|c|c|c|c}
\toprule
\multirow{2}{*}{Target Class} & \multirow{2}{*}{Classifier} & \multicolumn{5}{c|}{Task A: Random Train-Test-Split} & \multicolumn{5}{|c}{Task B: LOSO-CV}\\\cline{3-12}
 &  &  F1 Score & Acc. & Precision & Recall & R. to Bln. &  F1 Score & Acc. & Precision & Recall &  R. to Bln.\\
\hline
\multirow{3}{*}{Tap Water} 
  & Baseline       & 0.11 & 0.50 & 0.06 & 0.49 & -- & 0.11 & 0.50 & 0.07 & 0.51 & --      \\
  & CNN            & \textbf{0.75} & 0.97 & 0.75 & \textbf{0.76} & \textbf{668.29}\,\% & \textbf{0.48} & 0.92 & 0.48 & \textbf{0.54} & \textbf{446.83\,\%} \\
  & Random Forest  & 0.74 & \textbf{0.97} & \textbf{0.92} & 0.62 & 655.68\,\% & 0.47 & \textbf{0.95} & \textbf{0.76} & 0.35 & 427.39\,\% \\
\midrule
\multirow{3}{*}{Water}
  & Baseline       & 0.17 & 0.50 & 0.10 & 0.50 & -- & 0.17 & 0.50 & 0.10 & 0.51 & --     \\
  & CNN            & \textbf{0.71} & 0.94 & 0.73 & \textbf{0.69} & \textbf{409.63\,\%} & \textbf{0.52} & 0.89 & 0.51 & \textbf{0.60} & \textbf{314.56\,\%}  \\
  & Random Forest  & 0.68 & \textbf{0.95} & \textbf{0.92} & 0.53 & 391.29\,\% & 0.45 & \textbf{0.92} & \textbf{0.78} & 0.33 & 280.98\,\% \\

\bottomrule
\end{tabular}

\end{table*}